# Percolation model for the superconductor-insulator transition in granular films

Yakov M. Strelniker, A. Frydman, and S. Havlin

*Minerva Center, Jack and Pearl Resnick Institute of Advanced Technology,*
*and Department of Physic, Bar-Ilan University,*
*IL-52900 Ramat-Gan, Israel*

(Dated: December 24, 2006)

We study the temperature dependence of the superconductor-insulator transition in granular superconductors. Empirically, these systems are characterized by very broad resistance tails, which depend exponentially on the temperature, and the normal state resistance. We model these systems by a two-dimensional random resistor percolation networks in which the resistance between two grains is governed either by Josephson junction coupling or by quasi particle tunneling. Our numerical simulations as well as an effective medium evaluation explain the experimental results over a wide range of temperatures and resistances. Using effective medium approximation we find an analytical expression for the effective resistance of the system and the value of the critical resistance separating conducting from insulating branches.

PACS numbers: 74.25.Fy,74.50.+r,74.81.Fa,77.84.Lf,74.78.Na, 64.60.Ak, 02.50.-r, 73.23.-b

The disorder driven superconductor-insulator transition (SIT) in thin films has gained revived attention over the past few years mainly because of the possibility that it represents a quantum phase transition[1–6]. Disordered superconductors can be categorically divided into two groups, granular and homogeneous[9]. Experimentally, homogeneous samples are found to exhibit sharp superconducting transitions and the crossover between insulating to superconducting behavior seems to occur at $R \cong h/4e^2 \approx 6.5$ k$\Omega$.[1] On the other hand, granular films are characterized by very broad tails in the resistance $R(T)$ and the transition between the insulator and the superconductor phases seems to be much less universal.[2–8] In these samples it was found that the temperature dependence of the sheet resistance, $R(T)$, below the critical temperature, $T_c$, can be described by an inverse Arrhenius law:

$$R(T) = \mathcal{R}_0 \exp(T/T_0), \qquad (1)$$

where $T_0$ and $\mathcal{R}_0$ are constants[5]. The $R(T)$ curves following Eq. (1) have been observed for a large variety of granular superconductors and over a wide range of temperatures (down to temperatures below $T_c/10$). A typical example for a set of measurements[5] of $R(T)$ for quench condensed Pb granular films is shown in Fig. 1(a).

In this Letter we present a model based on percolation to account for the observed temperature dependence of the resistance in granular superconductors.[10] Our percolation approach is based on a two-dimensional (2D) random disordered array of grains, each neighboring pair represent a superconducting junction in which transport can be achieved either by Josephson tunneling or by quasi-particle tunneling, depending on the inter-grain coupling and temperature.[11] Different thickness of the film is represented in our percolation model by different distributions of inter-grain distances. Our numerical simulations of such a system exhibit an exponential dependence of $R(T)$ over a large range of temperatures which is in good agreement with recent experiments. We find that the critical resistance that separates insulating from superconducting branches depends on the distribution of disorder and on the nature of the percolation network of the current trajectories[12,13].

We begin by considering a 2D array of lattice sites, each one representing a superconducting grain. If two grains are sufficiently decoupled, the conductivity between two neighboring sites, $i$ and $j$, is given by[14–16]:

$$\sigma_{ij} = (e^2 \gamma_{ij}^0/k_B T) \exp(-r_{ij}/r_0 - \epsilon_{ij}/k_B T), \qquad (2)$$

where $\gamma_{ij}^0$ is a rate constant related to the electron-phonon interaction[16] (usually of the order $10^{12}$ s$^{-1}$), $r_{ij}$ is the distance between the two sites, $r_0$ is the scale over which the wave-function decays outside the grain and $\epsilon_{ij} = (|E_i| + |E_j| + |E_i - E_j|)/2$ is the zero field activation energy, which can be determined from physical principals. In the case of superconducting grains this is a nontrivial problem, but in general $\epsilon_{ij}$ is related to the superconducting gap $\Delta_{ij}(T)$ and Coulomb energy $E_c \sim e^2/2C$ (where $C$ is capacitance and $e$ is the electron charge)[11,17]: $\epsilon_{ij} = \Delta_{ij}(T) + E_{c,ij}$. Since the grains are assumed to be large enough to sustain bulk superconductivity, we assume $\Delta(T)$ is the same for all grains.

To perform numerical simulations of this model, we assume that the random distance between grains is $r_{ij} = 2\bar{l} \cdot \xi_{ij}$, where $\xi_{ij}$ is a random number taken from a uniform distribution in the range (0,1), and $\bar{l}$ is the mean distance between metallic grains.[18] Therefore the term $r_{ij}/r_0$ can be expressed as $\kappa \xi_{ij}$, where $\kappa \equiv 2\bar{l}/r_0$ can be interpreted as the dimensionless mean hopping distance or as the degree of disorder[19] (lower density of the deposited grains represent larger $\kappa$). Similarly, the charging energy, $E_c$, can be expressed through the same factor $\kappa \xi_{ij}$:

$$E_c = \beta \frac{2e^2}{4\pi\epsilon_0 \epsilon d} \frac{r_{ij}}{d} \equiv \beta E_c^{(0)} \frac{r_0}{d} \kappa \xi_{ij}, \qquad (3)$$

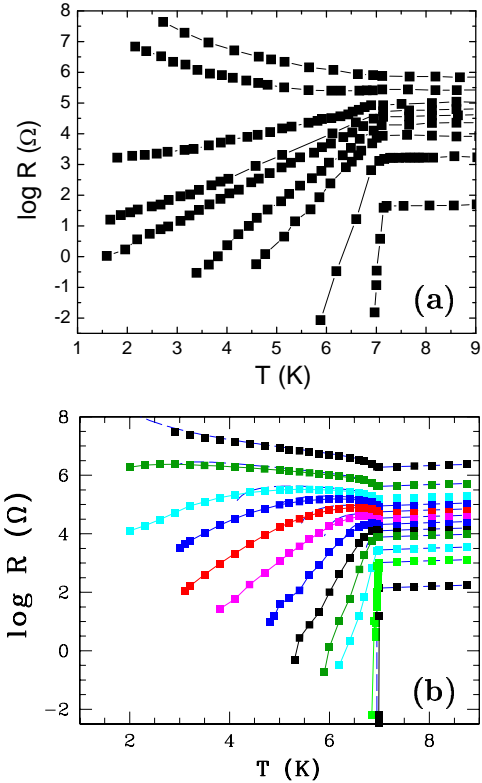

FIG. 1: (a) Experimental plots of the sheet resistance versus temperature for quench condensed Pb films[5] with grain sizes of the order $\sim 5 - 10$ nm, having different inter-grain coupling. The higher curves correspond to the thinner films, i.e., to films with larger inter-grain distance and, therefore, to the stronger disorder which is determined by $\kappa$). (b) (Color online) Theoretical plots of SIT: $\log R$ vs. $T$ for the systems with different disorder strength [see Eq. (4)], $\kappa$=25,22,20,19,18,17,16,15,14,12,10,6 (from top to bottom), respectively. Squares are the results of our numerical simulations, while the dashed lines are the plots of our analytical prediction (9). Sample size in simulations is $40 \times 40$ resistors, $T_c = 7$, $\alpha = 3.25 \cdot 10^3$, $z = 10$, and $R_0$ is proportional to temperature coefficient determined by Eq. (2).

where $d$, $s$, and $\epsilon$ are the size of the grains, the spacing between neighboring grains, and the dielectric constant, of the substrate, respectively[20,22]. The value $E_c^{(0)} = 2e^2/4\pi\epsilon_0\epsilon d$ is a charging energy of a single grain, and $\beta \sim 0.1$ is the effective coefficient which was invoked as a result of the influence of the surrounding grains.[20]

Finally we can write the expression for the local net resistor mimicking the local hopping resistance between the grains $R_{ij} = 1/\sigma_{ij}$ in the convenient form:

$$R_{ij} = R_0 \exp[\tilde{\kappa}\xi_{ij} + \Delta(T)/k_B T], \qquad (4)$$

where $\tilde{\kappa} \equiv \kappa(1 + \mathcal{E}_c/T)$, $\mathcal{E}_c \equiv \beta E_c^{(0)}(r_0/d)/k_B$, $R_0$ is a coefficient, proportional to the temperature, $T$, and determined by Eq. (2).

The superconducting gap $\Delta(T)$ is the solution of the integral equation[23,24] $\ln(\Delta(0)/\Delta) = 2I[\Delta(T)/T]$, where $I(u) \equiv \int_0^\infty [\sqrt{x^2 + u^2}(\exp\sqrt{x^2 + u^2} + 1)]^{-1} dx$. For temperatures near the critical value $T \sim T_c$, the gap $\Delta(T)$ can be approximated by the analytical form[23,25] $\Delta(T) = 3.06 k_B T_c (1 - T/T_c)^{1/2}$.

For each pair of grains the Josephson binding energy is given by[11,24,26,27]:

$$E_J = \alpha \left[\Delta(T)/R_{ij}^{(n)}\right] \tanh\left[\Delta(T)/2k_B T\right], \qquad (5)$$

where $\alpha = \pi\hbar/4e^2 \cong 6.5/2 = 3.25$ k$\Omega$ and $R_{ij}^{(n)} = R_{ij} \exp[-\Delta(T)/k_B T]$, is the normal resistance between the grains. The Josephson energy $E_J$ is related to the Josephson current $J_J$ as follows: $E_J = (\hbar/2e)J_J$. The criterion for the two neighboring grains to become "phase locked"[20,21] is

$$zE_J \geq k_B T + E_c, \qquad (6)$$

where $E_c$ is the energy of Coulomb blockade and $z$ is a parameter of the order of the number of the nearest-neighboring grains.[11,20] Since the Coulomb energy can be expressed through the parameter $\kappa$ [see Eq. (3)], we can write a self-consistency equation for the upper value of $R_{ij}^{(n)}$ for which Eq. (6) is fulfilled,

$$R_{JC} = \frac{z\alpha\delta(T)\tanh[\delta(T)/2]}{1 + \frac{\mathcal{E}_c}{T}\frac{\ln(R_{JC}/R_0)}{(1+\mathcal{E}_c/T)}}. \qquad (7)$$

$R_{JC}$ is the local critical parameter. When $R_{ij}^{(n)} \leq R_{JC}$ the neighboring $i$-th and $j$-th grains are Josephson coupled. Here we have used the relation $\kappa\xi_{ij} = \ln(R_{ij}^{(n)}/R_0)/(1+\mathcal{E}_c/T)$, and introduced definition $\delta(T) = \Delta(T)/k_B T$. In the case of small $E_c$ (classical SIT)[28], inequality (6) can be reduced to the well known condition[11,26,27]: $R_{ij}^{(n)} \leq z\alpha\delta(T)\tanh[\delta(T)/2]$.

Next we aim to evaluate the total resistance of the network system. Our numerical simulations were performed considering a 2D bond-percolating resistor network where $R_{ij}$ of each resistor is zero if Eq. (6) is fulfilled, otherwise it is given by Eq. (4). We solve the obtained system of linear Kirchhoff equations[18,29] and calculate the total effective resistance, $R(T)$, of the 2D network. The results are shown in Fig. 1(b). These results are in good agreement with the experimental data shown in Fig. 1(a).[30] From Fig. 1 we can also see, that the SIT is a result of an interplay between quasi-particle tunneling, which tends to turn the curves up [i.e., to increase resistance, $R$, with decreasing the temperature, $T$, see Eq. (4)] and Josephson coupling mechanism, which tend to turn curves down [i.e., to decrease the resistance, $R$, due to increase the total number of the Josephson junctions, which is proportional to superconducting gap $\Delta(T)$, see Eqs. (5)-(7)]. To qualitatively understand the behavior obtained in our simulations, which are very similar to the experiments (Fig. 1), we describe the percolation mechanism leading to this behavior. For large $\kappa$ (strong disorder), from percolation theory follows, that actually the current flows along a single path on the percolation cluster which is the path with minimal total resistance[31] and the total resistance of the path is determined by few critical (red bonds[32]) resistors[31,33]. The tail decrease of $R$ in Fig. 1 can be understood since when $T$ decreases (below $T_c$), $R_{JC}$ increases and more of these critical resistors become superconductors.

It is seen that an inverse Arrhenius dependence of $R(T)$ is obtained over a wide range of temperatures. Our agreement with experiments is further demonstrated in Fig. 2 which shows the dependence of $T_0$ [the slope of the tails, see Eq. (1)] as a function of the normal state resistance of the sample, $R_N$, for both experimental and simulation results.

For further understanding of this complex transition, we have also calculated the resistance of such a network using the symmetric self-consistency effective medium approximation (EMA)[18,34]. The effective resistance $R_e$ of the random conductance network [the local resistivities of which are distributed continuously in a range $R_{\min} \leq R \leq R_{\max}$ according to some distribution function $f(R)$], can be found as a solution of the integral equation

$$\int_{R_{\min}}^{R_{\max}} \theta(Re^{-\frac{\Delta}{k_BT}} - R_{JC})f(R)\left(\frac{R-R_e}{aR+R_e}\right)dR$$
$$-\int_{R_{\min}}^{R_{\max}} \theta(R_{JC} - Re^{-\frac{\Delta}{k_BT}})f(R)dR = 0, \qquad (8)$$

where $R_{JC}$ is determined by Eq. (7) (subscripts $ij$ have been omitted), $a = z/2 - 1$ and $z$ is the number of bonds at each node of the network. Here we split the integral into two parts in order to take the Josephson coupling into account in accordance with the condition (7), and have used the fact that $R^{(n)} = R\exp[-\Delta(T)/k_BT]$ [see Eq. (4)]. In the first integral we calculate the cases in which $R$ is larger than necessary for the Josephson coupling $R > R_{JC}$ [i.e., when $\theta(R - R_{JC}) = 1$, where $\theta$ is the Heaviside function]. In the second integral we consider the opposite situation, i.e., when $\theta(R_{JC} - R) = 1$. In this case Josephson coupling exists and $R$ in the brackets should taken as zero ($R \to 0$). If $\xi$ in Eq. (4) is uniformly distributed between 0 and 1, then $f(R) = 1/\kappa R$, and $R$ is varied in the range $R_0e^{\delta} \leq R \leq R_0e^{\delta+\kappa}$, see Eq. (4).

From Eq. (8) we can find the solution (for $R_0 < R_{JC} \leq R_0e^{\delta+\kappa}$)

$$R_e = \frac{1-p_c}{p_c}\frac{(R_0e^{\tilde{\kappa}p_c} - R_{JC})e^{\Delta/k_BT}}{1-e^{-\tilde{\kappa}(1-p_c)}}, \qquad (9)$$

where $R_{JC}$ is given by Eq. (7), and $p_c = 1/(1+a) = 2/z$ (see also Refs. 18,34).

Eq. (9) can be understood qualitatively as follows: The total resistance of the system at $T > T_c$ [when $\Delta(T) =$

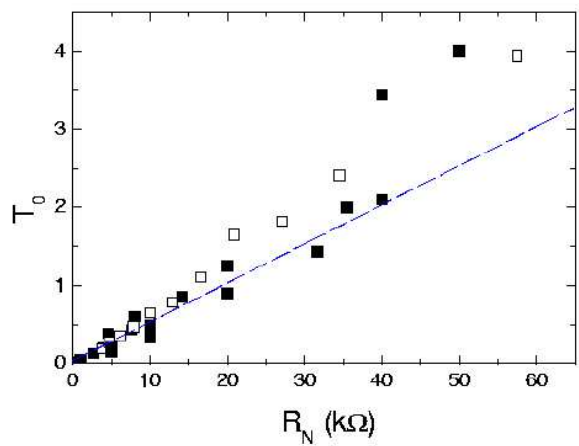

FIG. 2: The inverse slopes of the $R(T)$ tails [$T_0$ of Eq. (1)] as a function of the normal state resistance $R_N$[5] for several sets of Pb granular films measurements[5] (full squares) and our simulation results (empty squares). The dashed line is our analytical prediction, Eq. (10).

0] and for large $\kappa$ is equal to $R_0 e^{p_c\kappa+[p_c\beta E_c+\Delta(T)]/k_BT}$ (when $z = 4$, i.e., $p_c = 0.5$). As discussed above, if the system is strongly disordered, then its total resistance is determined by few resistors[32,33] along a path on the spanning cluster[16,18,19]. At $T < T_C$ same of the grains along this path have $R_{ij}^{(n)}$ smaller than $R_{JC}$ and, according to Eq. (7), these will be in the superconducting state. Therefore this resistance, proportional to $R_{JC}$, should be subtracted from the total resistance: $R_e = (R_0 e^{p_c\tilde{\kappa}} - R_{JC})e^{\Delta/k_BT}$.

We expand Eq. (9) linearly near the critical temperature ($T \sim T_c$), from which we get an expression for $R_e$ linear in terms $(T_c - T)$: $R_e \simeq R_0 e^{\delta(T)+p_c\kappa} - (\tilde{\alpha}/T_c)(T_c - T)$, where $\tilde{\alpha} = 3.06^2 z\alpha/2$ (see also Ref. 27). In the same approximation we get $\ln[R_e/R_N] \simeq -\tilde{\alpha}e^{-p_c\kappa}(T_c - T)/T_c$, where $R_N \equiv R_e(T_c) = R_0\exp(p_c\kappa)$. Note, that the inverse Arrhenius law (1) follows immediately from the latter expression with

$$T_0 = (T_c/\tilde{\alpha})R_N, \qquad (10)$$

shown in Fig. 2 in comparison to experimental and numerical results. It should be noted, that the system with exponential disorder (2) behaves differently than the system with two levels of resistivity (e.g., $\rho = 0$, and $\rho = 1$).

The analytical results can be also used in order to find the critical value of the effective resistance $R_{cr}^{(e)}$ separating the metal-like ($\partial R_e/\partial T \geq 0$), and insulator-like ($\partial R_e/\partial T < 0$) behaviors. By taking the derivative and solving the equation $\partial R_e/\partial T = 0$, we get a self-consistency equation, which determines the critical value separating metal-like behavior (at $R \leq R_{cr}$) from insulator like (at $R > R_{cr}$). The value of $R_{cr}$ [as well as $R_e$, see Eq. (9)] depends on two main factors: number of neighboring grains $z$ and charging energy $\mathcal{E}_c$. For small $\mathcal{E}_c$ we get a simple expression

$$R_{cr} = z\alpha(1-p_c)/p_c. \qquad (11)$$

In summary we have modeled and studied the temperature dependence of the superconductor-insulator transition in granular superconductors. Our numerical simulations explain well the experimental results over a wide range of temperatures and resistances. Calculations of effective medium approximation also show excellent agreement with the experiments for temperatures close to $T_c$. These calculations also enable us to determine the critical resistance value, separating the superconducting and insulating branches.

This research was supported by grants from the US-Israel Binational Science Foundation, the Israel Science Foundation, the European research NEST Project No. DYSONET 012911, and the KAMEA Fellowship program of the Ministry of Absorption of the State of Israel. We thank Profs. V. Sandomirsky, B. Shapiro, I. Shlimak, D. Stauffer, and P. Sheng for valuable conversations.